\documentclass{elsart}
\usepackage{graphicx}
\begin{document}
\begin{frontmatter}
\title{On the Magnetic-Field Dependence of \\ the Longitudinal Ultrasonic 
	  Attenuation \\ in a Type-II Superconductor}
\author{Kazue Kudo}
\ead{kudo@degway.phys.ocha.ac.jp}
\address{Department of Physics, Faculty of Science 
	      and Graduate School of Humanities and Sciences, 
	      Ochanomizu University, 2-1-1 Otsuka, Bunkyo-ku, Tokyo
	  112-8610, Japan }

\begin{abstract}
We propose a simple method by which we can explain the magnetic-field dependence of
the longitudinal ultrasonic attenuation in a type-II
superconductor. It gives a curve which is in good agreement with
 experimental data, in paticular, near the lower critical field
 $H_{c1}$. We compare it with conventional methods, which is not in
 good agreement with the experimental data near $H_{c1}$ but near the
 upper critical field $H_{c2}$. 
\end{abstract}

\end{frontmatter}
\section{Introduction}

The temperature dependence of the ultrasonic attenuation in a
superconductor is well described by the BCS theory~\cite{BCS}.
According to the BCS theory, the longitudinal
ultrasonic attenuation coefficient in the superconducting state
relative to that in the normal state is given by
\begin{equation}
 \frac{\alpha_{\mbox{s}}}{\alpha_{\mbox{n}}} = \frac2{\exp\left[ \frac{ {\Delta}
(T)}{k_B T}\right] +1} .
\label{eqn:1}
\end{equation}
The energy gap parameter ${\Delta}(T)$ depends on the temperature.

However, the magnetic-field dependence has not been explained very
well. Ikushima {\it et al.} performed an experiment about the
magnetic-field dependence of the longitudinal ultrasonic attenuation
in the mixed state of pure niobium~\cite{ikushima}. And they explained 
 it theoretically. The spatial average of the energy gap,
$\langle {\Delta}(H,T) \rangle$, was introduced and assumed
to be proportional to the root mean square of the order
parameter. Then $\langle {\Delta}(H,T) \rangle$ must be
proportional to the square root of the magnetization, $M$, near the
upper critical field, $H_{c2}$. They assumed this relation to be valid
throughout the mixed state. And they calculated the attenuation
coefficient in the mixed state, $\alpha_{\mbox{s}}$, with the measured
magnetization. The results of calculation give the curve which is
not in good agreement with the experimental data except near $H_{c2}$.

In this paper, we propose an original simple method  to explain the 
magnetic-field
dependence of the longitudinal ultrasonic attenuation, and compare it
with conventional methods. First, we study the original method,
 which is to estimate the 
ratio of the space ocuppied with vortices. It gives a curve which is 
in good agreement
with the experimental data, in paticular, near $H_{c1}$. We also study the conventional
method, which is to calculate the order parameter by solving the Ginzburg-Landau
(GL) equations. It gives a curve which is not in good agreement
with experimental data near $H_{c1}$ but near $H_{c2}$.
In each method, we use eq.~(\ref{eqn:1}), in which 
$\langle {\Delta}(H,T) \rangle$ is used instead of 
$\langle {\Delta}(T) \rangle$.

\section{To estimate the ratio of the space occupied with vortices}

\begin{figure}
 \begin{center}
   \includegraphics[height=8cm]{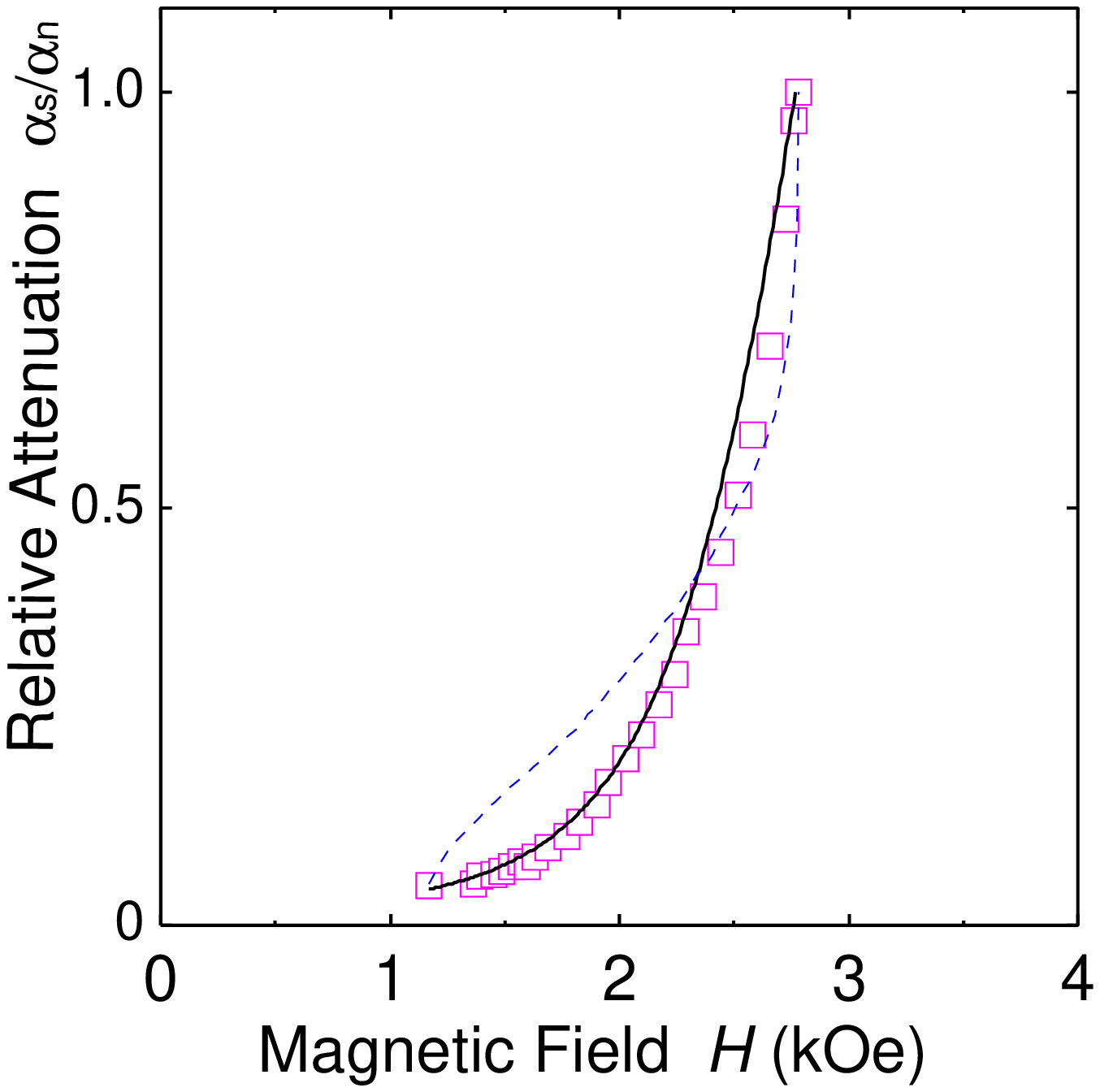} 
 \end{center}
   \setlength{\unitlength}{1cm}
\caption{\label{fig:1}
The longitudinal ultrasonic attenuation coefficient in the
 mixed state relative to that in the normal state. It is plotted against
 the applied magnetic field $H$ at 4.2K. The lower end of the solid
 curve corresponds to $H_{c1}$, and the upper end to $H_{c2}$. The open
 squares, experimental data, and the dashed curve, theoretical curve,
 are from ref.\protect\cite{ikushima}. The solid curve is
 obtained from eqs.~(\protect\ref{eqn:1}) and (\protect\ref{eqn:7}).}
\end{figure}

We take $\eta$ to be the ratio of the space ocuppied with
vortices. For example, if each vortex is assumed to be a cylinder
whose radius is the coherence length $\xi$, $\eta$ can be written
as
\begin{equation}
\eta = n\pi \xi^2 .
\label{eqn:2}
\end{equation}
Here, $n$ is the number of vortices per unit area on the plane
perpendicular to the magnetic field. We should notice that not only
$n$ but also $\xi$ varies with the field~\cite{sonier,golubov}. Let us
assume that the energy gap vanishes in each vortex and equals
${\Delta}(T)$ in the other area. Then we can introduce the
following equation:
\begin{equation}
\langle {\Delta}(H,T) \rangle = (1-\eta )
{\Delta}(T).
\label{eqn:3}
\end{equation}
Next, let us introduce a simple and phenomenological assumption:
the increase in the magnetic-field energy equals that of the free
energy when the field increases from $H_{c1}$ to $H$:
\begin{equation}
\frac1{8\pi} ( H^2-H_{c1}^2 ) =
\Omega_{\mbox{s}} (H)-\Omega_{\mbox{s}} (H_{c1}),
\label{eqn:4}
\end{equation}
and the increase of the free energy should be related to
$\eta$. Thus we have
\begin{equation}
\eta = A ( H^2-H_{c1}^2 ).
\label{eqn:5}
\end{equation}
Here, A is a constant. Taking into account that 
$\langle {\Delta}(H,T) \rangle$ vanishes at
$H=H_{c2}$, from eqs.~(\ref{eqn:3}) and (\ref{eqn:5}) we have
\begin{equation}
A= \frac1{H_{c2}^2-H_{c1}^2} .
\label{eqn:6}
\end{equation}
Combining eqs.~(\ref{eqn:3}), (\ref{eqn:5}) and (\ref{eqn:6}), we
obtain 
\begin{equation}
\langle {\Delta}(H,T) \rangle=
\left( 1- \frac{H^2-H_{c1}^2}{H_{c2}^2-H_{c1}^2} \right)
{\Delta}(T).
\label{eqn:7}
\end{equation}

The solid curve in Fig.~\ref{fig:1} is obtained from
eqs.~(\ref{eqn:1}) and (\ref{eqn:7}). It is in good agreement with the
experimental data, particularly near $H_{c1}$.

\section {To calculate the order parameter by solving the GL equations}

Now, we calculate $\langle | {\Psi} |^2 \rangle$. Here, ${\Psi}$ is the
order parameter: ${\Psi} ={\Psi}_o f e^{i\gamma} $; 
${\Psi}_o$ is the order parameter in the absence  of field; $f$ and $\gamma$
are the normalized magnitude and phase of the order parameter. Hao
{\it et al.}~\cite{hao} solved the Ginzburg-Landau equations with a
trial function,
\begin{equation}
f = \frac{r}{(r^2+\xi_v^2)^{1/2}} f_{\infty}.
\label{eqn:8}
\end{equation}
Here, $\xi_v$ and $f_{\infty}$ are two variational parameters; $r$
is distance from a vortex axis. And, especially
for $\kappa \simeq 5$, the parameters were approximated by the
following formulas:
\begin{eqnarray}
f_{\infty}^2 = 1- \left( \frac{B}{\kappa} \right)^4, \\
\label{eqn:9}
\left( \frac{\xi_v}{\xi_{vo}} \right)^2 = 
1+ \left( \frac{B}{\kappa} \right)^4,
\label{eqn:10}
\end{eqnarray}
where $B= 2\pi / (\kappa A_{\rm cell})$ is the averaged magnetic flux
density; $A_{\rm cell}$ is the unit cell area of the vortex lattice;
$\xi_{vo}$ is the value of $\xi_v$ at $B=0$. Using eqs.~(\ref{eqn:8}),
and (9) and (\ref{eqn:10}), the order parameter should be given by
\begin{equation}
 \frac{\langle | {\Psi} |^2 \rangle}{{\Psi}_o^2} =
\frac1{A_{\rm cell}} \int\!\!\!\int f^2 dS =
\frac1{A_{\rm cell}} \int\!\!\!\int
\frac{f_{\infty}^2 r^2}{r^2+\xi_v^2} dS.
\label{eqn:11}
\end{equation}
Here, the integral is taken over one lattice cell, which is
approximated by a circle centered at a vortex axis and has the same
cell area. By calculating eq.~(\ref{eqn:11}), we find
\begin{equation}
 \frac{\langle | {\Psi} |^2 \rangle}{{\Psi}_o^2} =f_{\infty}^2
\left[ 1- \frac{\xi_v^2}{R^2} 
\log\left(\frac{R^2}{\xi_v^2}+1\right) \right],
\label{eqn:12}
\end{equation}
where $R$ is the radius of the circle,
\begin{equation}
R^2=\frac{A_{cell}}{\pi} = \frac2{B\kappa}.
\label{eqn:13}
\end{equation}
And, in order to write 
$\langle | {\Psi} |^2 \rangle$ as a function of $H$, 
we assume that $B$ is related to $H$ by
the following equation~\cite{hao},
\begin{eqnarray}
H & = & \frac{\kappa f_{\infty}^2 \xi_v^2}2 \left[
\frac{1- f_{\infty}^2}2 
\ln \left( \frac2{B \kappa \xi_v^2} +1 \right)
- \frac{1-f_{\infty}^2}{2+B \kappa \xi_v^2} 
+ \frac{f_{\infty}^2}{(2+B \kappa \xi_v^2)^2} \right] \nonumber \\
& + & \frac{f_{\infty}^2(2+3B\kappa\xi_v^2)} 
{2\kappa (2+B\kappa\xi_v^2)^3}
+B +
\frac{f_{\infty}}{2\kappa\xi_v K_1(f_{\infty} \xi_v)} \nonumber \\
& \times & \left[
K_0\left( \xi_v (f_{\infty}^2+2B\kappa)^{1/2} \right)
-\frac{B\kappa\xi_v K_1\left( \xi_v(f_{\infty}^2+2B\kappa)^{1/2} \right)}
{(f_{\infty}^2+2B\kappa)^{1/2}} \right],
\label{eqn:14}
\end{eqnarray}
Then, if we use the relation,
\begin{equation}
\frac{\langle {\Delta}(H,T) \rangle }{ {\Delta}(T)} =
\sqrt{ \frac{\langle | {\Psi} |^2 \rangle}{{\Psi}_o^2} }.
\label{eqn:15}
\end{equation}
\begin{figure}
 \begin{center}
   \includegraphics[height=8cm]{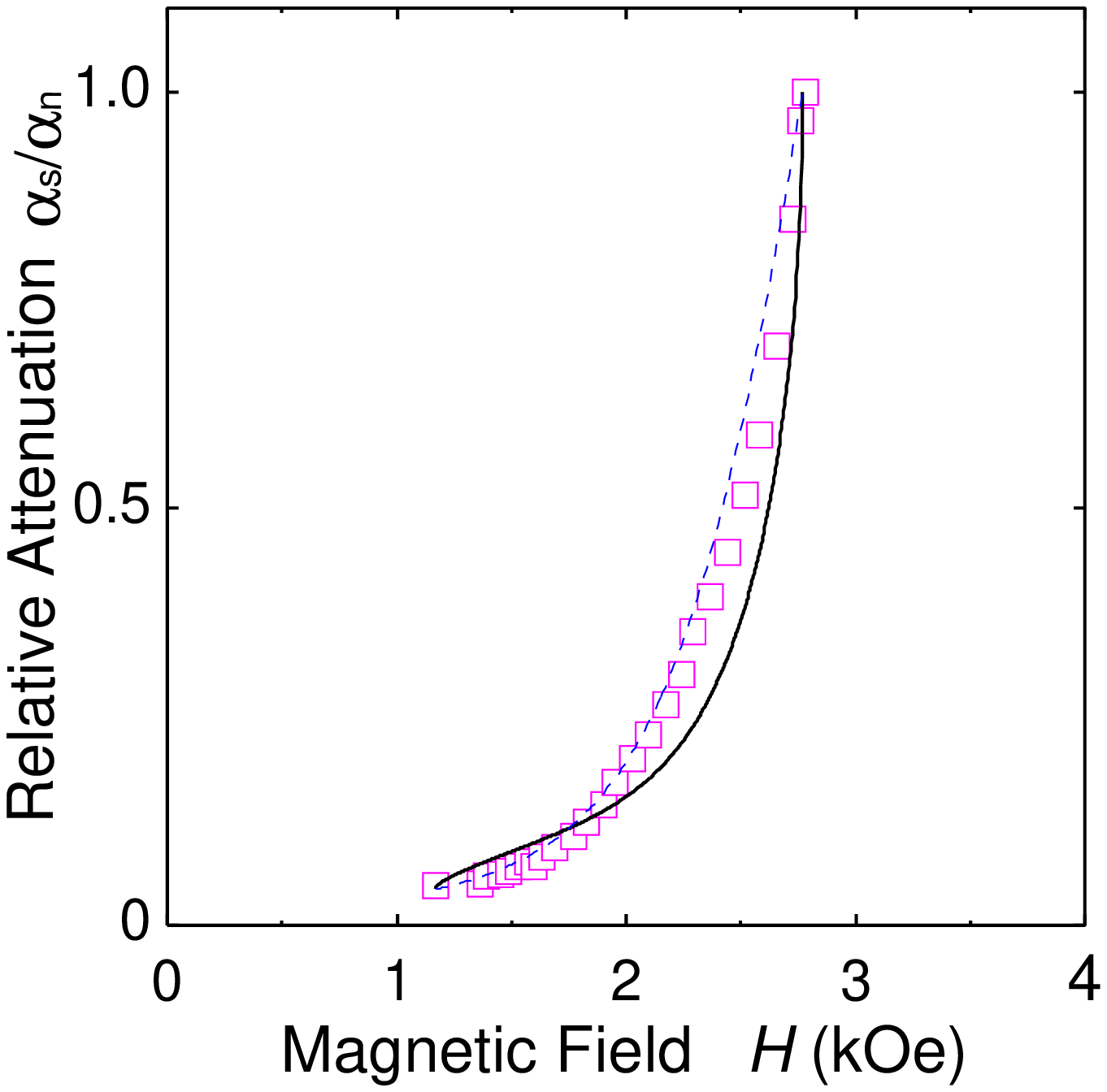} 
 \end{center}
\caption{\label{fig:2}The longitudinal ultrasonic attenuation coefficient in the
 mixed state relative to that in the normal state. The solid curve is
 obtained from eqs.~(\ref{eqn:1}) and (9)-(\ref{eqn:15}). The
 dashed curve is obtained from eps.~(\protect\ref{eqn:1}) and 
(\protect\ref{eqn:7}). The open squares are
 experimental data from ref.~\protect\cite{ikushima}. }
\end{figure}
The solid curve in  Fig.~\ref{fig:2} is obtained from eqs.~(\ref{eqn:1}) 
and (9)-(\ref{eqn:15}). It is not in good agreement with the experimental
data near $H_{c1}$ but near $H_{c2}$.

\section{Discussion}

In this section, we discuss the validity for each of the approximations 
which are
adopted to calculate the three theoretical curves.

The theoretical curve calculated by  Ikushima {\it et al.}~\cite{ikushima}
  was proposed to explain the experimental data particularly near $H_{c2}$.
 They used the approximation that $\langle {\Delta}(H,T) \rangle$ should
  be 
  proportional to the square root of the magnetization $M$. It is valid
  near $H_{c2}$. However, it is not necessarily valid near $H_{c1}$. 

In Section 2, we introduced simple and phenomenological assumptions and
approximations. Equations (4) and (5) should be effective near
$H_{c1}$, where the number of the vortices is small. However, they may be
less effective near $H_{c2}$. Because the relation between the free
energy and $\eta$, the ratio of the space occupied with vortices, may
be more
complex when the number of the vortices is large.
 
The approximation leading to eqs.~(9) and (\ref{eqn:10}) in Section 3 is not
necessarily  
effective in explaining $\langle {\Delta}(H,T) \rangle$, on which 
the ultrasonic attenuation depends. It was originally adopted to explain the
magnetization curves in ref.~\cite{hao}. The magnetization should be
strongly dependent on the penetrarion depth $\lambda$. Also, the order
parameter $\Psi$ in Section 3 should be dependent on $\lambda$. On the
other hand, $\langle {\Delta}(H,T) \rangle$ should be dependent on the
coherence length $\xi$.
In eq.~(15), we assumed the behavior of $\langle {\Delta}(H,T) \rangle$
to be equivalent to that of $\langle | {\Psi} | \rangle$. In
other words, we assumed the magnetic-field dependence
of $\lambda$ to be equivalent to that of $\xi$.  
However, the magnetic-field
dependence of $\lambda$ is generaly different from that of
$\xi$~\cite{sonier}. Therefore the magnetic-field dependence of 
magnetization should be
generaly different from that of $\langle {\Delta}(H,T) \rangle$. Equations (9) 
and (\ref{eqn:10}) may be effective in explaining 
$\langle {\Delta}(H,T) \rangle$ not through the whole mixed-state region
but only near $H_{c2}$.

\section{Conclusion}

The original method discussed in this paper gives the theoretical 
curve which is in good
agreement with experimental data particularly near $H_{c1}$. On the other
hand, the theoretical curves given by the two conventional methods are not
in good agreement with experimental data near $H_{c1}$ but $H_{c2}$.
Though the original method is simple and phenomenological, it may give 
some information on the conformation of the vortices 
in type-II 
superconductors particularly near $H_{c1}$.

\section*{ Acknowledgement.}

The author wishes to thank Dr. T.~Deguchi of Ochanomizu University for
his advice.

\end{document}